\documentclass[aps,pop,twocolumn,superscriptaddress,letterpaper]{revtex4}
\usepackage{mathrsfs}
\usepackage{amsmath}
\usepackage{bm}
\usepackage{color}
\usepackage{graphicx}
\usepackage{hyperref}
\usepackage{pdfpages}
\usepackage{pgffor}

\begin{document}


\title{Developing a Linear Fluid Plasma Model with Accurate Kinetic Bernstein Waves: A First Step}

\author{Huasheng Xie}
\email[]{Email: huashengxie@gmail.com, xiehuasheng@enn.cn} 
\affiliation{Hebei Key Laboratory of Compact Fusion, Langfang 065001, China}
\affiliation{ENN Science and Technology Development Co., Ltd., Langfang 065001, China}

\date{\today}

\begin{abstract}
Kinetic models provide highly accurate descriptions of plasma waves but involve complex integrals that are computationally expensive to solve. To facilitate a fluid-like treatment of the system, we propose rational approximations for both the plasma dispersion function in the parallel integral and the Bessel function in the perpendicular integral, ensuring that the system remains rational with respect to all three variables: wave frequency $\omega$, parallel wavevector $k_\parallel$, and perpendicular wavevector $k_\perp$. By accurately approximating the Bessel function over a wide range of Larmor radius $\rho_{cs}$ values, from $k_\perp\rho_{cs} \to 0$ to $k_\perp\rho_{cs} \to \infty$, we present an initial attempt to incorporate kinetic Bernstein waves into a fluid model. As an application, we employ this model to analyze { electromagnetic plasma} wave propagation conditions (i.e., accessibility) by solving for the complex perpendicular wavevector $k_\perp$ using a matrix eigenvalue method with given input parameters. This work may contribute to studies of electron cyclotron resonance heating (ECRH) and ion cyclotron resonance frequency (ICRF) heating in magnetized confinement plasmas.
\end{abstract}


\maketitle

\begin{figure*}
\centering
\includegraphics[width=18cm]{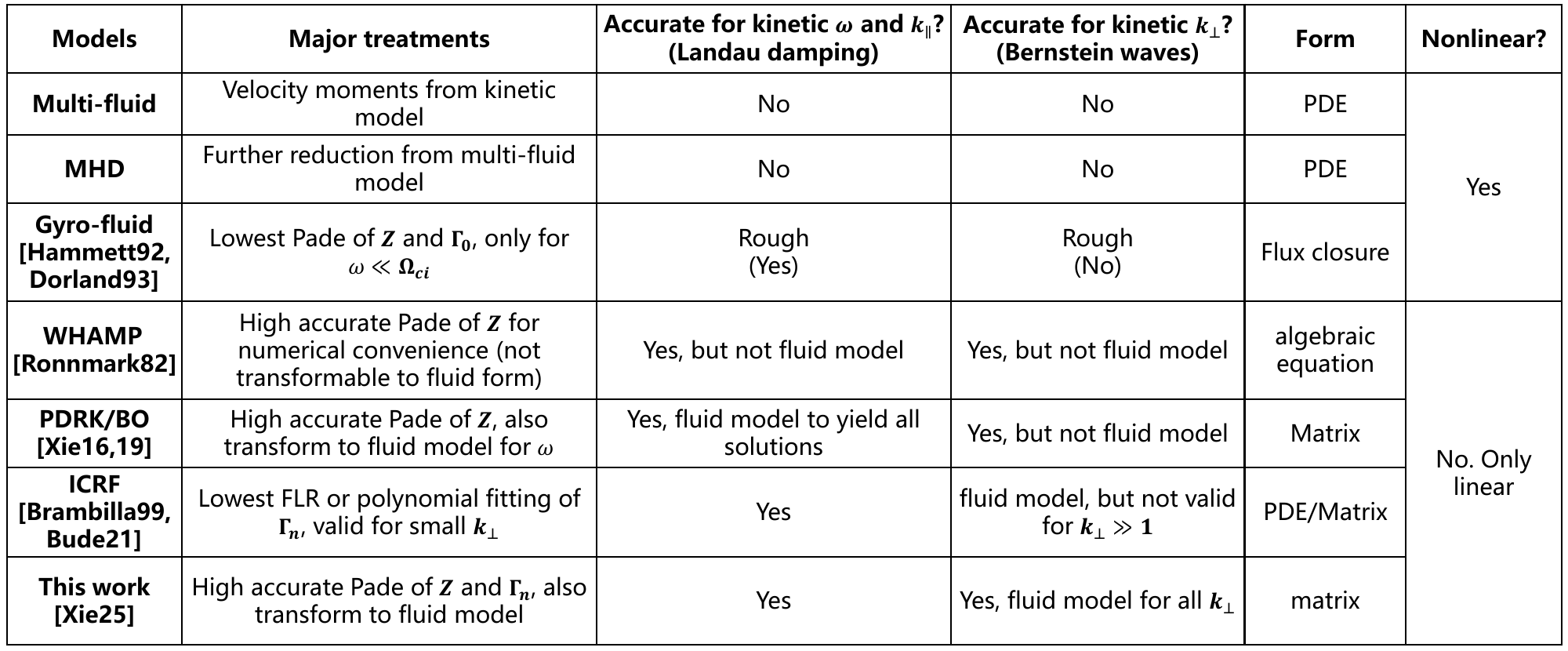}\\
\caption{{Comparison of fluid-relevant plasma models. This work initiates a fluid framework that accurately incorporates Bernstein modes and Landau damping. Nonlinear models are typically PDE-based (partial differential equation), while linear models can adopt matrix formulations.}}\label{fig:model_cmp}
\end{figure*}

{ 
Plasmas exhibit complex collective behavior arising from interactions between charged particles and electromagnetic fields, manifesting as waves, instabilities, and nonlinear transport processes. These phenomena are most accurately described using kinetic models, which account for the velocity-space distribution function and its evolution. In magnetized plasmas, two of the most fundamental kinetic effects are Landau damping~\cite{Landau1946} and Bernstein modes~\cite{Bernstein1958}. Landau damping represents collisionless wave dissipation due to resonant interactions between waves and particles, while Bernstein modes correspond to wave branches that exist at all harmonic orders of the cyclotron frequency—not just the fundamental.

Although kinetic models provide high physical fidelity, they are computationally intensive due to the requirement of evaluating velocity-space integrals. In contrast, fluid models dramatically reduce computational cost by taking velocity moments of the kinetic equation, yielding multi-fluid and magnetohydrodynamic (MHD) formulations. However, these conventional fluid models inherently fail to capture key kinetic effects such as Landau damping and Bernstein modes.

It would be highly desirable to develop a computationally efficient, fluid-like model that can still capture these essential kinetic features. However, this remains a significant challenge. The gyro-Landau-fluid (GLF) model~\cite{Hammett1990,Hammett1992,Dorland1993,Held2020} represents one such attempt: it incorporates Landau damping approximately via Padé expansions of the plasma dispersion function. While useful in studies of plasma turbulence and transport, the GLF model lacks the accuracy required for detailed analysis of linear wave propagation and does not account for Bernstein modes.

When the focus is restricted to linear wave physics, a more specialized approach is possible. In this context, we define a `fluid-like' model as one that avoids explicit velocity-space integration and instead expresses the system using rational functions of the wave frequency $\omega$, the parallel wavevector $k_\parallel$, and the perpendicular wavevector $k_\perp$. Such a formulation enables algebraic manipulation and allows the problem to be cast into a matrix form suitable for efficient numerical solution.

Figure~\ref{fig:model_cmp} compares representative fluid-related models, highlighting their respective capabilities and limitations. Among these, the accurate fluid treatment of $k_\perp$-dependent effects—crucial for describing Bernstein modes—remains an open challenge. The present work addresses this issue directly, and to our knowledge, provides the first fluid-like model capable of simultaneously capturing both Landau damping and Bernstein wave physics in a unified and algebraically tractable framework.

}

{ To illustrate the model’s applicability}, we consider local, linear, electromagnetic kinetic plasma waves in an infinite, homogeneous system, assuming a Maxwellian velocity distribution for each plasma species, given by  $f_{s0}={\pi^{-3/2}v_{ts}^{-3}} \exp[-(v_\parallel^2+v_\perp^2)/{v_{ts}^2}]$. The background magnetic field is assumed to be $ \bm{B}_0 = (0,0,B_0) $, and the wave vector is given by  
$ \bm{k} = (k_x,0,k_z) = (k\sin\theta,0,k\cos\theta) $,  
such that $ k_\parallel = k_z $ and $ k_\perp = k_x $. We consider $ S $ species, indexed by $ s = 1,2,\dots,S $. The electric charge, mass, density, and temperature of each species are denoted as $ q_s $, $ m_s $, $ n_{s0} $, and $ T_{s0} $, respectively. The thermal velocity is defined as  
$ v_{ts} = \sqrt{\frac{2k_B T_{s0}}{m_s}} $  
(some authors may use  $ v_{ts} = \sqrt{\frac{k_B T_{s0}}{m_s}} $),  
where $ k_B $ is the Boltzmann constant.

The non-relativistic kinetic dispersion relation (KDR) is
\begin{eqnarray}\label{eq:drkinetic}
  {D}(\omega,{\bm k})=|{\bm K}(\omega,{\bm k})+({\bm k}{\bm k}-k^2{\bm I})\frac{c^2}{\omega^2}|=0,
\end{eqnarray}
where
\begin{eqnarray}
  {\bm K}={\bm I}+{\bm Q}={\bm I}-\frac{{\bm \sigma}}{i\omega\epsilon_0},~~{\bm Q}=-\frac{{\bm \sigma}}{i\omega\epsilon_0}.
\end{eqnarray}

Define $a_s=k_\perp\rho_{cs}, ~b_s=a_s^2=k_\perp^2\rho_{cs}^2,~\rho_{cs}={v_{ts}}/{(\sqrt{2}\omega_{cs})},~
{\bm n}={{\bm k}c}/{\omega},~\omega_{cs}={q_sB_0}/{m_s},~\omega_{ps}=\sqrt{{n_{s0}q_s^2}/{(\epsilon_0m_s})},~c={1}/{\sqrt{\mu_0\epsilon_0}}$
and $\zeta_{sn}=({\omega-n\omega_{cs}})/{(k_zv_{ts})}$, where $c$ is the speed of light, $\omega_{ps}$ and $\omega_{cs}$ are the plasma frequency and cyclotron frequency of each species, $\epsilon_0$ is the permittivity of free space, and $\mu_0$ is the permeability of free space. Note that for electrons, $q_s < 0$, and thus $\omega_{cs}$, $\rho_{cs}$, and $a_s$ are negative. After standard derivations \cite{Stix1992,Brambilla1998,Xie2019}, we obtain
\begin{eqnarray}
  {\bm K}={\bm I}+\sum_s\frac{\omega_{ps}^2}{\omega^2}\Big[\sum_{n=-\infty}^{\infty}\zeta_{s0}Z(\zeta_{sn}){\bm X}_{sn}+2\zeta_{s0}^2{{\hat{\bm z}\hat{\bm z}}}\Big],
\end{eqnarray}
and
\begin{eqnarray}\nonumber
  {\bm X}_{sn} =\left[ {\begin{array}{ccc}
  \frac{n^2}{b_s}\Gamma_{sn} & in\Gamma'_{sn} & \sqrt{2}\zeta_{sn}\frac{n}{a_s}\Gamma_{sn}\\
  -in\Gamma'_{sn} & \frac{n^2\Gamma_{sn}}{b_s}-2b_s\Gamma'_{sn} & -i\sqrt{2}\zeta_{sn}a_s\Gamma'_{sn}\\
  \sqrt{2}\zeta_{sn}\frac{n}{a_s}\Gamma_{sn} & i\sqrt{2}\zeta_{sn}a_s\Gamma'_{sn} & 2\zeta_{sn}^2\Gamma_{sn} \end{array}}\right],
\end{eqnarray}
where $Z(\zeta)=\frac{1}{\sqrt{\pi}}\int_{-\infty}^{+\infty}\frac{e^{-z^2}}{z-\zeta}dz$ is the plasma dispersion function\cite{Xie2024}, $\Gamma_{sn}\equiv\Gamma_n(b_s)$ and  $\Gamma'_{sn}\equiv\Gamma'_n(b_s)$, with
$\Gamma_n(b)=I_n(b)e^{-b},~\Gamma'_n(b)=(I'_n-I_n)e^{-b},~ I'_n(b)=({I_{n+1}+I_{n-1}})/{2},~~I_{-n}=I_{n},~ Z'(\zeta)=-2[1+\zeta Z(\zeta)]$, and $I_n$ is the $n$-th order modified Bessel function. Note also the symmetry relations $K_{yx}=-K_{xy}$, $K_{zx}=K_{xz}$ and $K_{zy}=-K_{yz}$.

Due to the complexity of the plasma dispersion function $Z(\zeta)$ and the infinite-order summation of $\Gamma_n$, the KDR takes a complicated form in terms of $\omega$, $k_z$, and $k_x$. In particular, the integral form of $Z(\zeta)$ makes computations expensive. The search for a fluid model that can effectively mimic both the kinetic Landau damping effects from $Z(\zeta)$ and the finite Larmor radius (FLR) effects from the zeroth-order Bessel function term $\Gamma_0(b)$ has led to the widely used {gyro-Landau-fluid model \cite{Hammett1990,Hammett1992,Dorland1993,Held2020}}. However, while the gyro-Landau-fluid model is mainly used for nonlinear and non-uniform studies, it provides only a rough approximation of kinetic effects and is valid only for low-frequency waves, where $\omega \ll \omega_{ci}$.  

We are interested in capturing linear kinetic effects accurately within a fluid model, which is crucial for studying plasma waves. By a `fluid model', we mean a system that takes a simple algebraic form in terms of $\omega$, $k_z$, and $k_x$, or equivalently, in terms of $\partial / \partial t$, $\nabla_\parallel$, and $\nabla_\perp$. {This goal is achievable by the following two steps:  (1) Constructing accurate rational approximations for the plasma dispersion function $Z(\zeta)$ and the Bessel-related term $\Gamma_n(b)$ using a finite number of terms;  (2) Reformulating the resulting expressions into a structure analogous to fluid equations.  For the second step, we transform the system into an equivalent, but higher-dimensional, matrix form. This allows the dispersion relation to be expressed and solved as a generalized eigenvalue problem, while preserving key kinetic effects in an algebraic framework.
}

In the cold plasma limit or only for lowest order FLR effects, the approximations $\zeta \to \infty$ and $b \to 0$ are often used to simplify the model. However, we aim to construct a fluid model that remains valid across all ranges of $\zeta$ (which depends on $\omega$ and $k_z$) and $b$ (which depends on $k_x$).  

As a first step, we seek to construct an equivalent fluid model that accurately captures the major physical solutions of the KDR given by Eq.~(\ref{eq:drkinetic}). This is achieved through highly accurate rational approximations of $Z(\zeta)$ and $\Gamma_n(b)$, followed by a linear transformation into a matrix form. We present the details of this approach in this work.  

The function $Z(\zeta)$ has been approximated with high accuracy using Padé approximations and J-pole expansions \cite{Ronnmark1982,Ronnmark1983,Xie2024}, expressed as  $Z(\zeta)\simeq\sum_{j=1}^{J}\frac{b_j}{\zeta-c_j}$, where even for $J=8$, the error is less than $10^{-6}$. This method has been successfully applied to obtain all solutions of the KDR for $\omega$ using a matrix method \cite{Xie2016,Xie2019}, which remains valid for all significant solutions except for strongly damped modes, which are typically of lesser interest.

The accurate rational approximation of $\Gamma_n(b_s)$ is slightly complicated. For ion cyclotron resonance frequency (ICRF) studies, a Taylor expansion up to $\mathcal{O}(b_s)$ is commonly used \cite{Brambilla1999}. Additionally, a fitting approach using a truncated polynomial \cite{Bude2021,Eester2021} has been applied for $b_s > 1$. However, this method remains invalid as $b_s \to \infty$.  
Since our goal is to provide a rational form for all $\omega$, $k_z$, and $k_x$ in Eq.~(\ref{eq:drkinetic}), the formulation by Rönnmark \cite{Ronnmark1983} serves as a suitable starting point, where the plasma dispersion function $Z(\zeta)$ is already expressed in a rational form. In this approach, the term ${\bm Q}$ in the KDR is modified to  
\begin{eqnarray}\label{eq:Q}
  {\bm Q}&=&\sum_s\frac{\omega_{ps}^2}{\omega\omega_{cs}}\sum_jb_j\cdot\\\nonumber
&&\left[ {\begin{array}{ccc}
  R_{sj} & \frac{i}{x_{sj}}R'_{sj} & \frac{\sqrt{2}a_sc_j}{x_{sj}}R_{sj}\\
  -\frac{i}{x_{sj}}R'_{sj} & R_{sj}-\frac{2b_s}{x_{sj}^2}R'_{sj} & -\frac{i\sqrt{2}a_sc_j}{x_{sj}^2}R'_{sj}\\
  \frac{\sqrt{2}a_sc_j}{x_{sj}}R_{sj} & \frac{i\sqrt{2}a_sc_j}{x_{sj}^2}R'_{sj} & \frac{2c_j^2}{x_{sj}^2}(x_{sj}+b_sR_{sj}) \end{array}}\right],
\end{eqnarray}
where $b_j$ and $c_j$ are Padé approximation coefficients \cite{Ronnmark1982,Xie2024} for the $Z(\zeta)$ function, and $x_{sj} = (\omega - k_z v_{ts} c_j) / \omega_{cs}$ with $R_{sj} = R(x_{sj},b_s)$.  
Here{\small
\begin{eqnarray}
  R(x,\lambda)\equiv\sum_{n=-\infty}^{\infty}\frac{n^2\Gamma_n(\lambda)}{\lambda(x-n)}=-\frac{x}{\lambda}+x^2\sum_{n=-\infty}^{\infty}\frac{\Gamma_n(\lambda)}{\lambda(x-n)},\\
R'(x,\lambda)\equiv\frac{\partial [\lambda R(x,\lambda)]}{\partial \lambda}=\sum_{n=-\infty}^{\infty}\frac{n^2\Gamma'_n(\lambda)}{(x-n)}=x^2\sum_{n=-\infty}^{\infty}\frac{\Gamma'_n(\lambda)}{(x-n)},
\end{eqnarray}}
where $\sum_{n=-\infty}^{\infty}\Gamma_n(b)=1$, $\sum_{n=-\infty}^{\infty}n\Gamma_n(b)=0$ and $\sum_{n=-\infty}^{\infty}n^2\Gamma_n(b)=b$. There are two approaches to computing $R(x, \lambda)$. One method is to truncate the summation over $n$ to a finite $N$, which may be inaccurate for larger $\lambda$. Alternatively, a more precise approach is to use methods that account for all $n$, such as those discussed in \cite{Ronnmark1982,Schmitt1974}, ensuring accuracy for all values of $\lambda$.

\begin{figure}
\centering
\includegraphics[width=8.5cm]{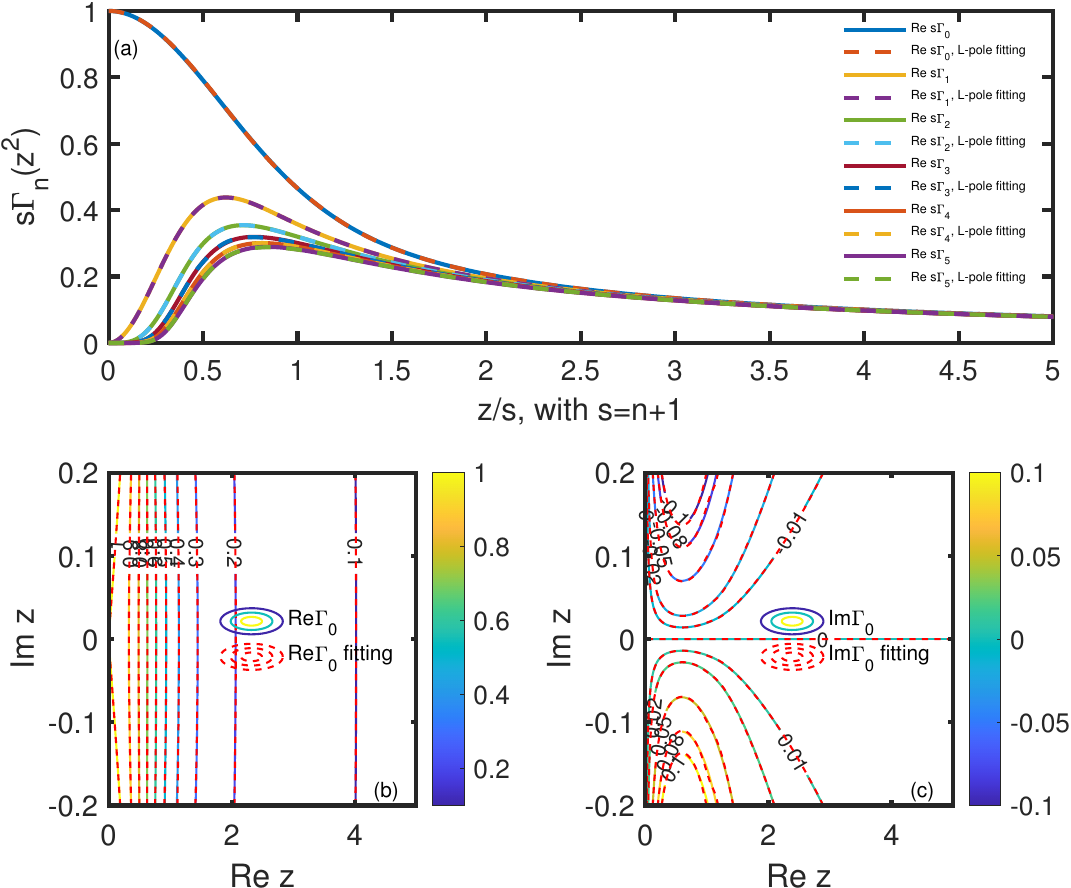}\\
\caption{The comparison of L-pole fitting for $\Gamma_n(z^2)$ {shows} excellent agreement for both the real and imaginary parts, even for ${\rm Im}(z)=0.2$.}\label{fig:cmp_GamnLpole}
\end{figure}

In this work, we fit $\Gamma_n(b)$ by truncating the summation $\sum_{n=-\infty}^{\infty}$ to $\sum_{n=-N}^{N}$. Considering that for large $b$ \cite{Abramowitz1972}, we have  
\begin{eqnarray}\nonumber
  \Gamma_n(b) &=& \frac{1}{\sqrt{2\pi b}} \left[ 1 - \frac{(4n^2 - 1)}{8b} + \right. \\\nonumber
  && \left. \frac{(4n^2 - 1)(4n^2 - 9)}{2!(8b)^2} - \cdots \right], \quad |\arg b| < \frac{\pi}{2},
\end{eqnarray}
and for small $b$,  
\begin{eqnarray}
  \Gamma_n(b) = e^{-b} \sum_{m=0}^{\infty} \frac{1}{m!\Gamma(m+n+1)} \left( \frac{b}{2} \right)^{2m+n},
\end{eqnarray}
with the expansion $e^{-b} = 1 - b + \frac{b^2}{2!} - \cdots$, where $\Gamma$ denotes the Euler gamma function.  
To approximate $\Gamma_n(b)$ in a rational form, we use an $L$-pole expansion ($n \geq 0$),  
\begin{eqnarray}\label{eq:Gamnfit}
  \Gamma_n(z^2) = \frac{z^{2n} (a_0 + a_2z^2 + \cdots + a_{L-1}z^{L-1})}{1 + \sqrt{2\pi} a_{L-1} z^{2n+L}} = \sum_{l_n=1}^{2n+L} \frac{r_{l_n}}{z - p_{l_n}}.
\end{eqnarray}
This form avoids singularities for $z>0$ along the real axis and ensures the asymptotic behavior $\Gamma_n(z^2 \to \infty) \simeq 1/(\sqrt{2\pi} z)$. Additionally, we note that $\Gamma_{-n} = \Gamma_n$.  To prevent unphysical solutions in Eq.~(\ref{eq:Q}), we enforce constraints up to ${O}(z^3)$ and ${O}(1/z^3)$. Specifically, for all $n$, we impose  $ a_{L-2} = 0, ~a_{L-3} = -\frac{(4n^2 - 1) a_{L-1}}{8}$. For $n \leq 3$, we set  $a_0 = \frac{1}{\Gamma(n+1)}$, and for $n=0$, we further impose $a_2 = -a_0$.
Equation~(\ref{eq:Gamnfit}) is valid for ${\rm Re}(z) \geq 0$, i.e., ${\rm Re}(k_x \rho_{cs}) \geq 0$. Thus, in practical computations, we select only solutions with ${\rm Re}(k_x) \geq 0$ and set $z = k_x |\rho_{cs}|$ in Eq.~(\ref{eq:Gamnfit}).  
Similar to the $J$-pole expansion of the $Z$ function \cite{Ronnmark1982,Xie2024}, it follows that  $\sum_{l_n}r_{l_n}p_{l_n}^0=1/\sqrt{2\pi}$, $\sum_{l_n}r_{l_n}p_{l_n}^1=0$  and $\sum_{l_n}r_{l_n}p_{l_n}^2=-(4n^2-1)/(8\sqrt{2\pi})$. Additionally, we have  $\sum_{l_n}r_{l_n}/p_{l_n}=0~ (n\geq1)$ or $1~(n=0)$, and $\sum_{l_n}r_{l_n}p_{l_n}^2=0$.

Under the enforced constraints on the coefficients $a_{l_n}$ ($l_n = 0,2,\dots,a_{L-1}$), they are determined using least-squares fitting over the range $z \in [-0.5, 10] \times (n+1)$. To maintain an error below 1\%, we require approximately $L \simeq 20$ for $n \geq 10$. For $n = 0$, $L = 12$ is sufficient to achieve high accuracy, while for $n = 1$ to $3$, $L = 16$ is adequate. Fitting up to $n \leq 10$ is sufficient for most applications in electron cyclotron resonance heating (ECRH) and ion cyclotron range of frequencies (ICRF) studies.
Since the analytical properties of $\Gamma_n$ are preserved for both small and large $z$, and the fitting equation remains in a single analytic form with good continuity properties, it is also valid for weakly imaginary $z$. This feature is crucial for wave absorption studies.

Figure~\ref{fig:cmp_GamnLpole} shows a comparison between the fitted $\Gamma_n$ and the exact values, demonstrating good agreement for both the real and imaginary parts.
A set of typical coefficients for $\Gamma_0$ with $L = 12$ is:
$r_{2l}$=[0.2944+0.2999i,   -0.3171+0.1391i,   -0.3533-0.06956i,   0.5887-0.462i,  -0.01285+0.04219i, -0.0003267-0.0004249i], $p_{2l}$=[-1.2+0.3215i, -0.8784+0.8784i, 0.3215+1.2i, -0.3215+1.2i, 0.8784+0.8784i,    1.2+0.3215i], where the remaining terms satisfy $p_{2l-1}=p_{2l}^*$ and $r_{2l-1}=r_{2l}^*$, with $^*$ denoting complex conjugation.
  
Similar to the derivation of the matrix method for solving $\omega$ in the KDR \cite{Xie2016, Xie2019} and for $k_x$ in the multi-fluid model \cite{Xie2021, Xie2024a}, we require only Maxwell’s equations along with a relation between the perturbed current $\delta{\bm J}$ and the perturbed electric field $\delta{\bm E}$, given by $\delta{\bm J}={\bm \sigma}\cdot\delta{\bm E}$. Maxwell’s equations can then be written as
\begin{subequations} \label{eq:matkxEB}
\begin{eqnarray}
  &k_x\delta E_y = \omega\delta B_z,\\
&  k_x\delta E_z = \Big(\frac{k_z^2c^2}{\omega}-\omega\Big)\delta B_y-\frac{ik_z}{\omega\epsilon_0}\delta J_x,\\
&  k_x\delta B_y = -\frac{\omega}{c^2}\delta E_z-\frac{i}{c^2\epsilon_0}\delta J_z,\\
&  k_x\delta B_z = \Big(\frac{\omega}{c^2}-\frac{k_z^2}{\omega}\Big)\delta E_y+\frac{i}{c^2\epsilon_0}\delta J_y,
  \end{eqnarray}
\end{subequations}
and $\delta E_x=(k_zc^2/\omega)\delta B_y-i\delta J_x/(\omega\epsilon_0)$ and $\delta B_x=-(k_z/\omega)\delta E_y$. It can be shown that after the L-pole rational expansion of $\Gamma_n$, we obtain
\begin{eqnarray}\label{eq:sigma}
  {\bm \sigma}=\epsilon_0\sum_l\left[ {\begin{array}{ccc}
  \frac{b_{lxx}}{k_x-c_l} & \frac{b_{lxy}}{k_x-c_l} & \frac{b_{lxz}}{k_x-c_l}\\
  \frac{b_{lyx}}{k_x-c_l} & \frac{b_{lyy}}{k_x-c_l} & \frac{b_{lyz}}{k_x-c_l}\\
  \frac{b_{lzx}}{k_x-c_l} & \frac{b_{lzy}}{k_x-c_l} & \frac{b_{lzz}}{k_x-c_l} \end{array}}\right],
\end{eqnarray}
where $l$ is summation for $s$, $n$ and $l_n$, i.e., $\sum_l=\sum_s\sum_{n=0}^{N}\sum_{l_n}$. The coefficients are $c_l=p_{l_n}/|\rho_{cs}|$ and
\begin{eqnarray*}\label{eq:bl}
  b_{lxx}&=&-2i\frac{\omega_{ps}^2}{\omega_{cs}|\rho_{cs}|}\sum_jb_j\frac{r_{l_n}}{p_{l_n}^2}x_{sj}\frac{n^2}{x_{sj}^2-n^2},\\
  b_{lxy}&=&2\frac{\omega_{ps}^2}{\omega_{cs}|\rho_{cs}|}\sum_jb_jr_{l_n}\Big[\frac{1}{2}\frac{(n+1)^2}{x_{sj}^2-(n+1)^2}\\&&
-(n\geq1)\cdot\frac{n^2}{x_{sj}^2-n^2}+(n\geq2)\cdot\frac{1}{2}\frac{(n-1)^2}{x_{sj}^2-(n-1)^2}\Big],\\
  b_{lxz}&=&-2\sqrt{2}i\frac{\omega_{ps}^2}{\omega_{cs}\rho_{cs}}\sum_jb_jc_{j}\frac{r_{l_n}}{p_{l_n}}\frac{n^2}{x_{sj}^2-n^2},\\
  b_{lyy}&=&-2i\frac{\omega_{ps}^2}{\omega_{cs}|\rho_{cs}|}\sum_jb_j\Big\{\frac{r_{l_n}}{p_{l_n}^2}x_{sj}\frac{n^2}{x_{sj}^2-n^2}\\&&-\frac{{r_{l_n}}{p_{l_n}^2}}{x_{sj}}\Big[\frac{(n+1)^2}{x_{sj}^2-(n+1)^2}
-(n\geq1)\cdot2\frac{n^2}{x_{sj}^2-n^2}\\&&+(n\geq2)\cdot\frac{(n-1)^2}{x_{sj}^2-(n-1)^2}\Big]\Big\},\\
  b_{lyz}&=&-2\sqrt{2}\frac{\omega_{ps}^2}{\omega_{cs}\rho_{cs}}\sum_j\frac{b_jc_j}{x_{sj}}{r_{l_n}}{p_{l_n}}\Big[\frac{1}{2}\frac{(n+1)^2}{x_{sj}^2-(n+1)^2}\\&&
-(n\geq1)\cdot\frac{n^2}{x_{sj}^2-n^2}+(n\geq2)\cdot\frac{1}{2}\frac{(n-1)^2}{x_{sj}^2-(n-1)^2}\Big],\\
  b_{lzz}&=&-2i\frac{\omega_{ps}^2}{\omega_{cs}|\rho_{cs}|}\sum_jb_jc_j^2{r_{l_n}}x_{sj}\frac{[1+(n\geq1)]}{x_{sj}^2-n^2},
\end{eqnarray*}
and $b_{lyx}=-b_{lxy}$, $b_{lzx}=b_{lxz}$ and $b_{lzy}=-b_{lyz}$. The matrix form can readily be obtained as
\begin{subequations} \label{eq:matkxJ}
\begin{eqnarray}\nonumber
  k_x\delta v_{xl} &=& c_l\delta v_{xl}+b_{lxy}\delta E_y+b_{lxz}\delta E_z\\&&+b_{lxx}\frac{1}{\omega}\Big(k_zc^2\delta B_y-\frac{i\sum_{l'}\delta v_{xl'}}{\epsilon_0}\Big),\\\nonumber
  k_x\delta v_{yl} &=& c_l\delta v_{yl}+b_{lyy}\delta E_y+b_{lyz}\delta E_z\\&&+b_{lyx}\frac{1}{\omega}\Big(k_zc^2\delta B_y-\frac{i\sum_{l'}\delta v_{xl'}}{\epsilon_0}\Big),\\\nonumber
  k_x\delta v_{zl} &=& c_l\delta v_{zl}+b_{lzy}\delta E_y+b_{lzz}\delta E_z\\&&+b_{lzx}\frac{1}{\omega}\Big(k_zc^2\delta B_y-\frac{i\sum_{l'}\delta v_{xl'}}{\epsilon_0}\Big),
  \end{eqnarray}
\end{subequations}
where we have used $\delta J_x=\sum_l\delta v_{xl}$, $\delta J_y=\sum_l\delta v_{yl}$ and $\delta J_z=\sum_l\delta v_{zl}$.

\begin{figure*}
\centering
\includegraphics[width=18cm]{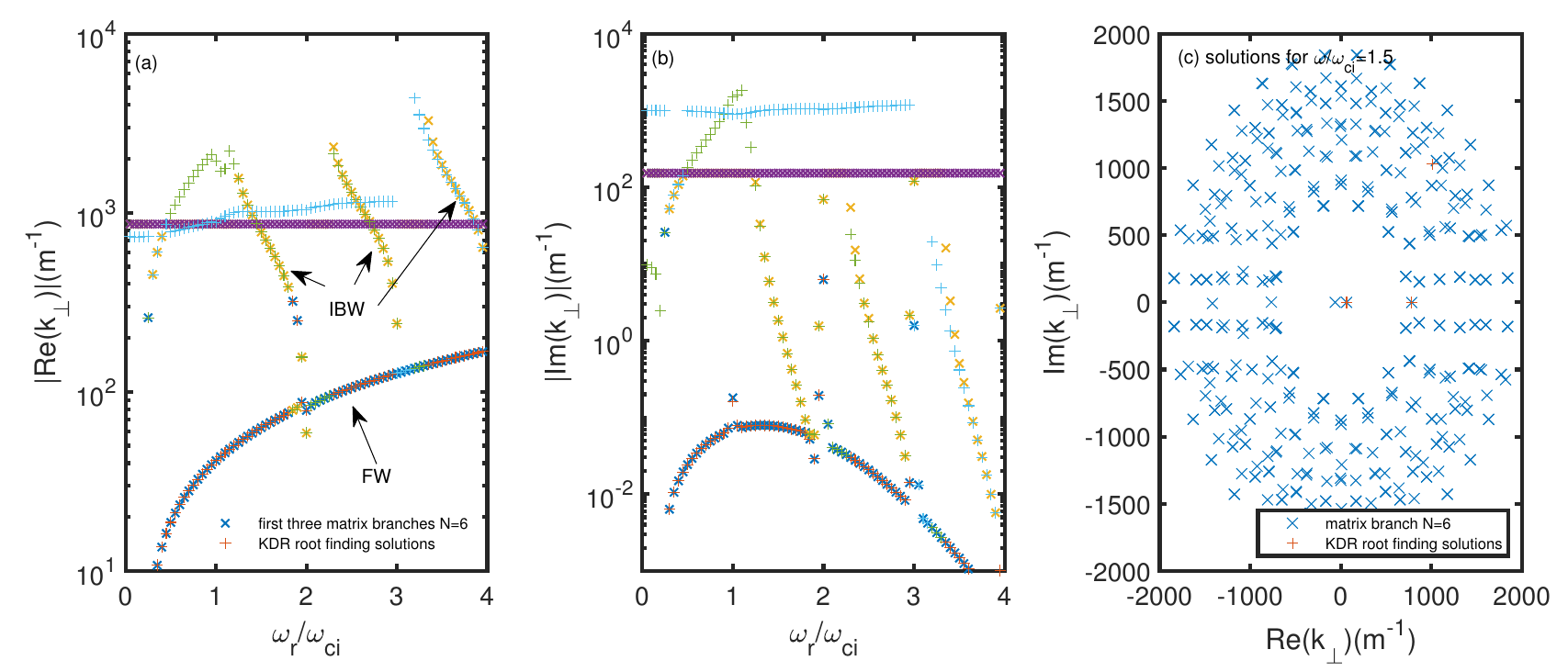}\\
\caption{The comparison of ICW in the fluid matrix system described by Eqs. (\ref{eq:matkxEB}) and (\ref{eq:matkxJ}) with KDR solutions shows excellent agreement for most major solutions, both for the real part ${\rm Re}(k_\perp)$ (a) and the imaginary part ${\rm Im}(k_\perp)$ (b). This indicates that the fluid model can accurately capture {ICW fast waves (FW) and IBW}. More solutions from the fluid matrix are shown in (c) for $\omega/\omega_{ci} = 1.5$.}\label{fig:cmp_icw}
\end{figure*}

\begin{figure*}
\centering
\includegraphics[width=18cm]{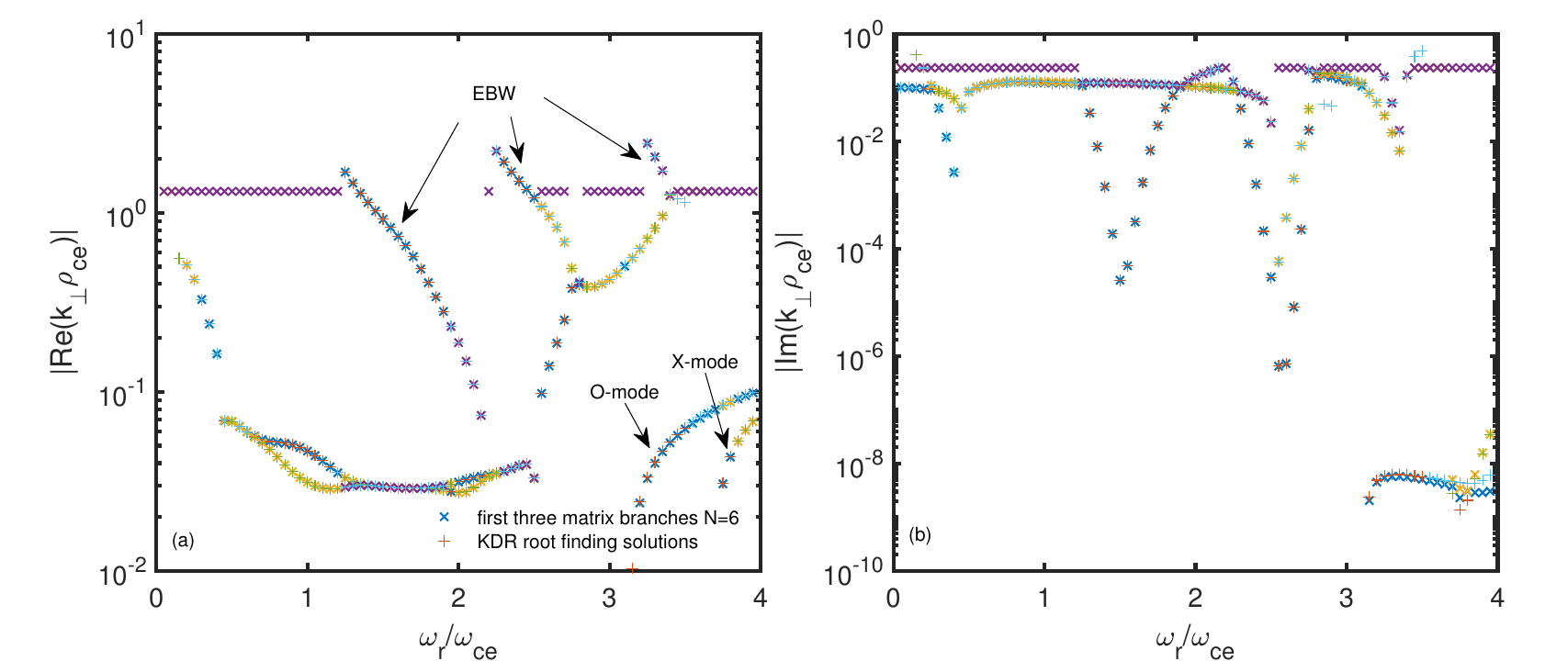}\\
\caption{The comparison of ECW in the fluid matrix system described by Eqs. (\ref{eq:matkxEB}) and (\ref{eq:matkxJ}) with KDR solutions shows good agreement for most major solutions, both for the real part ${\rm Re}(k_\perp)$ (a) and the imaginary part ${\rm Im}(k_\perp)$ (b). This {illustrates} that the fluid model can accurately capture {ECW O-mode and X-mode waves, as well as EBW}.}\label{fig:cmp_ecw}
\end{figure*}

The fluid model is constructed by combining Eqs.~(\ref{eq:matkxEB}) and (\ref{eq:matkxJ}). This linear system can be solved as a matrix eigenvalue problem, $k_x \mathbf{X} = \mathbf{M} \cdot \mathbf{X}$, to obtain all solutions of $k_x$, where the state vector is defined as ${\bm X}=[\delta v_{xl},\delta v_{yl},\delta v_{zl},\delta E_{y},\delta E_{z},\delta B_{y},\delta B_{z}]$.  
This model has a structure similar to the warm multi-fluid model~\cite{Xie2014, Xie2021, Xie2024a} but with significantly larger dimensions and complex-valued coefficients. The matrix dimension is approximately $S\times\sum_n{L_n}+4\simeq S\times N\times (L+N)+4$, whereas the matrix dimension of the warm multi-fluid model is only $2S+4$~\cite{Xie2024a}.  

For a complex frequency $\omega = \omega_r + i\omega_i$, the dominant mode is typically the one with the largest imaginary part $\omega_i$. However, for a complex wavenumber $k_x = k_{xr} + i k_{xi}$, the selection criterion differs. We consider the first few solutions with $k_{xr} > 0$ and the smallest $|k_{xi}|$ as the physically relevant solutions. This choice is justified because solutions with large \( k_{xi} \) (either rapidly growing or strongly damped) are typically non-physical due to boundary constraints and are unlikely to dominate in realistic scenarios.

{
In the cold plasma model, three main wave modes are identified within the cyclotron frequency regimes. In the electron cyclotron frequency range, $\omega \sim \omega_{ce}$, two electron cyclotron waves (ECWs) exist: the extraordinary mode (X-mode) and the ordinary mode (O-mode). In the ion cyclotron frequency range, typically $\omega \sim \omega_{ci}$, one ion cyclotron wave (ICW) appears, commonly referred to as the fast wave (FW). When kinetic effects are included, these cold-plasma wave branches are modified due to Landau damping and finite Larmor radius effects. This results in the emergence of new Bernstein branches, namely electron Bernstein waves (EBWs) and ion Bernstein waves (IBWs), which play a critical role in wave absorption and plasma heating.}

To {illustrate} the accuracy of the proposed fluid model, we compare its solutions with those of the kinetic dispersion relation (KDR) in both the ICW and ECW regimes. This comparison is particularly relevant for analyzing wave accessibility and propagation characteristics in magnetized plasmas. The parameters used in these tests are taken from Fig.~3 of~\cite{Xie2016} for the ICW case and Fig.~12 of~\cite{Xie2021} for the ECW case. The corresponding results are shown in Fig.~\ref{fig:cmp_icw} and Fig.~\ref{fig:cmp_ecw}, respectively.

The fluid matrix model exhibits excellent agreement with the KDR across a broad frequency range. It successfully reproduces the main features of ion Bernstein waves (IBWs) and electron Bernstein waves (EBWs), as well as ICW fast waves (FW), and ECW X-mode and O-mode solutions. In particular, the agreement is strong near the resonant condition $\omega_r - k_\parallel v_{ts} - n\omega_{cs} \simeq 0$, indicating that the model accurately captures wave absorption mechanisms. Minor discrepancies are observed only in the strongly damped solutions, which are typically less physically significant. These differences can be attributed to limitations in the $\Gamma_n$ fitting for large imaginary arguments and the truncation of the summation over harmonic numbers $n$ to a finite $N$.

The input parameters for the ICW case are: $S=2$ (electron and deuterium ion), $B_0=6$T, $n_{s0}=2e20{\rm m^{-3}}$, $T_s=4{\rm keV}$ and $k_z=10{\rm m^{-1}}$, yielding $\rho_{ci}=0.00152$m and $f_{ci}=\omega_{ci}/2\pi=45.7$MHz. The input parameters for the ECW case are: $S=1$ (electron only), $B_0=1$T, $n_{s0}=6.075e19{\rm m^{-3}}$, $T_s=1{\rm keV}$ and $k_z\rho_{ce}=0.1$, yielding  $\rho_{ce}=7.54e-5$m and $f_{ce}=\omega_{ci}/2\pi=28$GHz. The small differences in the imaginary part $k_{xi}$ between the ECW O- and X-mode solutions and the KDR results may be due to round-off fitting errors. Since the ratio $k_{xi}/k_{xr}$ is on the order of $10^{-6}$, this discrepancy is negligible.  

The strong agreement between the fluid model, given by Eqs.~(\ref{eq:matkxEB}) and (\ref{eq:matkxJ}), and the KDR results provides confidence that kinetic effects, such as Bernstein waves, are accurately incorporated into the fluid framework. This could be a crucial step toward developing further applications.  
For a fixed $x$, it is also possible to fit $R(x,\lambda)$ directly \cite{Ronnmark1982,Schmitt1974}, which would be useful for high-frequency harmonics requiring large $N$ to converge for large $\lambda$. However, this approach necessitates refitting each time $x$ changes, which depends on $\omega$, $k_z$ and $v_{ts}$. 

In summary, this study {presents the first successful construction of a fluid-like model that accurately captures both kinetic Landau damping and Bernstein modes, achieved} precise rational approximations and transformation into an equivalent matrix representation. The fluid matrix method avoids numerical difficulties such as divergence and the need for good initial values when solving the KDR. This approach can be directly applied to { plasma wave propagation conditions analysis \cite{Xie2021,Xie2024a} and} ray tracing studies \cite{Xie2022}, and we anticipate further developments in fast full-wave simulations of ICRF \cite{Bude2021,Zhang2024} using fluid models with accurate kinetic effects. {For further extensions, the Pad\'e-based arbitrary-wavelength polarization closures for gyrokinetic and gyrofluid models \cite{Held2020} may also provide a similar approach.} For ECW modeling, relativistic effects play a significant role and represent an important direction for future research.

The author would like to thank the discussions with Jiahui Zhang and Haojie Ma.  The source code for this work is available at \url{https://github.com/hsxie/fluidbw}.

\end{document}